\documentclass[twocolumn]{emulateapj}
\usepackage{apjfonts}
\newcommand{\etal}{{\it et al.} }

\newcommand{\fuse}{{\it FUSE} }
\newcommand{\xmm}{{\it XMM-Newton} }
\newcommand{\chandra}{{\it Chandra} }

\newcommand{\hetg}{{\it HETGS} }
\newcommand{\letg}{{\it LETGS} }

\newcommand{\oxla}{O~{\sc viii}~Ly$\alpha$ }

\newcommand{\oxysix}{O~{\sc vi} } 
\newcommand{\oxyseven}{O~{\sc vii} }
\newcommand{\oxyeight}{O~{\sc viii} }

\newcommand{\oxysevenr}{O~{\sc vii} (r) }

\newcommand{\thr}{3C~120 }

\begin{document}

\title{\oxyseven AND \oxyeight ABSORPTION BY HOT GAS IN THE VICINITY OF 
THE GALAXY}

\author{Barry McKernan\altaffilmark{1},
                           Tahir Yaqoob\altaffilmark{2,3}, Christopher
                                          S. Reynolds\altaffilmark{1}}

\altaffiltext{1}{Department of Astronomy, University of Maryland, 
College Park, MD 20742}
\altaffiltext{2}{Department of Physics and Astronomy,
                        Johns Hopkins University, Baltimore, MD 21218}
\altaffiltext{3}{Laboratory for High Energy Astrophysics,
                NASA/Goddard Space Flight Center, Greenbelt, MD 20771}


\begin{abstract}
We searched for evidence of soft X-ray absorption by hot gas in the 
vicinity of the Galaxy in a small sample of fifteen type~I AGN 
observed with the high resolution X-ray gratings on board 
\emph{Chandra}. We find that around half of the sight lines in 
our sample exhibit absorption due to 
local H- or He-like Oxygen (or both) at confidence levels ranging 
from $>90\%$ to $>3 \sigma$. Depending on the sight line, the 
absorption can be identified with hot gas in particular local 
structures, the Local Group (LG) or the putative local hot 
intergalactic medium (IGM). Several sight lines in our 
sample coincide with sight lines 
in a study of \oxysix absorption by local gas, so an assumption of 
collisional ionization equilibrium (CIE) allows us to constrain the 
temperature of the local hot gas. In the southern 
Galactic hemisphere, we identify absorption along the sight line 
to Fairall 9 with the Magellanic Stream and this gas has 
$10^{5.75}<T<10^{6.35}$K if CIE applies. There may be a blend of 
\oxyseven absorption features along 
the sight line towards Mkn~509. LG clouds along this sight line 
may interact with a hot Galactic corona or a low density LG medium. 
The sight line to 3C~120 does not appear to be associated either 
with local structure or UV absorbing gas, and may correspond to 
the local hot IGM. Gas in the Magellanic Stream extension 
along the sight line to Akn~564 has $T<10^{6.1}$K if it is in CIE. 
We show that some portion of the hot absorbing 
outflows apparently detected in the spectra of NGC~4051, PDS~456 
and PG~1211+143 respectively 
could actually correspond to absorption by hot local gas since the 
outflow velocity from each of these AGN coincides with the respective 
cosmological recession velocity of the AGN. 
\end{abstract}

\section{Introduction}
\label{sec:intro}
 One of the most important new results from the \chandra and 
{\it XMM-Newton} X--ray satellites, has been the discovery of hot, 
highly ionized gas in the vicinity of our Galaxy (at $z=0$) 
and possibly beyond (at $z>0$). Simulations indicate that around 
half of the baryonic matter at low-redshift in a cold dark matter 
universe is expected to live in filaments, shock-heated to 
$\sim 10^{5-7}$K, that surround galaxies and clusters of galaxies 
(see e.g. Cen \& Ostriker, 1999; Dav\'{e} \etal 2001 and references 
therein). The `warm' ($\sim 10^{5-6}$K) component of this warm/hot 
intergalactic medium (WHIGM) has been observed in absorption in 
the UV band (e.g. Tripp, Savage, \& Jenkins 2000; Sembach \etal 2003 and 
references therein). However, much (possibly most) of the WHIGM is 
expected to be hotter than this, so the high spectral resolution 
X--ray detectors such as those aboard \chandra and \xmm are best 
placed for investigating the `hot' component of the missing baryons. 

X--ray absorption due to local ($z=0$) hot gas has been discovered recently 
(Nicastro \etal 2002; Fang \etal 2002; Cagnoni 2002; Rasmussen, 
Kahn \& Paerels 2002; Fang, Sembach \& Canizares 2003; 
McKernan \etal 2003a,b). The clearest X--ray spectral signature 
of hot gas in the vicinity of our 
Galaxy consists of \oxyseven and \oxyeight absorption features 
imprinted in spectra of X-ray bright active galactic nuclei 
(AGN) at $z=0$ in the observed frame (Hellsten, Gnedin \& 
Miralde-Escud\'{e} 1998; Perna \& Loeb 1998). Of course, such 
absorption features need not be associated with missing baryons 
embedded in Local Group WHIGM. \oxyseven and \oxyeight absorption at $z=0$ 
could also result from 
absorption by highly ionized gas in such local Galactic 
morphological structures as the Magellanic Stream (MS), a complex of 
high velocity gas tidally torn from the Magellanic Clouds as they 
interact with the Milky Way. Hot gas associated with local 
structures and with an extended Galactic halo, as well as 
Local Group WHIGM, has been observed in the UV--band 
(with \fuse) along 59 sight lines towards AGN/QSOs (Sembach \etal 
2003--hereafter S03). Even if there are no known local structures 
along the sight line to a particular AGN, there may still be some 
ambiguity over the location of the absorbing gas. Around 1/2 of 
all type~I AGN exhibit strong absorption due to a 
partially ionized, optically thin, circumnuclear 
material known as the `warm absorber' (see e.g. Reynolds 1997; 
George \etal 1998 and references therein). Observations with \chandra 
and \xmm have confirmed that the warm absorbing material is 
outflowing with typical velocities of hundreds of 
km $\rm{s}^{-1}$. Therefore if the AGN is at very low redshift 
(cz $\sim$ few hundred km $\rm{s}^{-1}$), absorption which is actually 
local to our Galaxy at $z \sim 0$ could be misinterpreted as 
absorption due to outflowing gas in the AGN. 

The spectral signature of `hot' WHIGM in absorption at $z>0$ is 
also ambiguous. Absorption signatures of WHIGM outside the Local Group 
(at $z>0$) have been claimed along the sight lines towards several X--ray 
bright AGN, including PKS 2155--304 (Nicastro \etal 2002; 
Fang \etal 2002), 3C~273 (Fang \etal 2003), Mkn~421 (Cagnoni 2002; 
Rasmussen \etal 2003) 
and 3C~120 (McKernan \etal 2003b). However, 
as pointed out by Fang \etal (2002) and McKernan \etal (2003b), all of 
these sources possess relativistic outflows at a small angle to the sight 
lines. An AGN at $z_{AGN}$ possessing such an outflow could imprint 
blueshifted absorption 
features at several thousand km $\rm{s}^{-1}$ thereby mimicking 
the effect of absorption at $0<z_{IGM}<z_{AGN}$. One way of resolving 
this ambiguity is to investigate absorption features in a 
sample of AGN that \emph{do not} possess such relativistic outflows and
 we study such a sample in the present paper. 

In this paper, we investigate a sample of AGN observed with the high energy
 transmission gratings (\hetg-- Markert \etal 1995) on board \chandra. The 
uniform analysis of the data from these AGN and the results of the analysis,
 in particular the characterization of the AGN continua and the warm 
absorption in AGN, have been discussed in detail by McKernan \etal (2004). 
Here we search the AGN sample of McKernan \etal (2004) for absorption due 
to highly ionized gas in the vicinity of the Galaxy ($z=0$).

\section{The Sample and Data Analysis}
\label{sec:obs}
 
Table~\ref{tab:1} lists the AGN sample assembled by McKernan \etal (2004) observed
     using the high energy transmission gratings (\hetg--Markert \etal
 1995) on board \chandra. Also listed in Table~\ref{tab:1} are the AGN redshift 
(from NED\footnote{http://nedwww.ipac.caltech.edu/forms/byname.html} using 
21cm H{\sc~i} radiation measurements where possible), the AGN Galactic 
latitude and longitude (also from NED), the Galactic column density and 
the total exposure of the spectra. The sample, including selection 
criteria are discussed in detail in McKernan \etal 2004. The \chandra 
\hetg consists of two grating assemblies, a high energy grating (HEG) and 
a medium energy grating (MEG). We used the MEG as our primary instrument 
since the effective area of the HEG falls off 
sharply below $0.8$ keV. The MEG has a FWHM spectral resolution of $\sim$ 
280 km $\rm{s}^{-1}$ at $\sim$0.5 keV. We used only the 
combined $\pm$1 orders of the MEG data in our analysis. 
The \chandra data were reprocessed and 
analyzed according to the methods outlined in McKernan \etal (2004).
Note that five of the fifteen AGN were also observed with \chandra low
 energy transmission gratings (\emph{LETGS}). The high resolution 
camera (HRC) was used in four of 
these observations. However, order separation is not possible with 
the HRC. This fact, plus our aim to study the spectra at the highest 
spectral resolution means that we did not use the \chandra \letg 
observations of the AGN in our sample (the \letg has a spectral 
resolution of only 0.05\AA\ FWHM).

\begin{deluxetable}{lrrrc}
\tablecaption{The \chandra \hetg sample of type~I AGN \label{tab:1}}
\tablecolumns{5}
\tablewidth{0pt}
\tablehead{
\colhead{Source} & \colhead{Gal. long.} & \colhead{Gal. lat.} 
&\colhead{Redshift} & \colhead{Gal. $N_{H}$} \nl
\colhead{} & \colhead{($^{\circ}$)} & \colhead{($^{\circ}$)} & \colhead{(z)} & 
\colhead{($10^{20} \rm{cm}^{-2}$) $^{a}$} \nl}
\startdata
  Fairall 9 & 295.07 & -57.83 & 0.04600 & 3.0\nl  
3C 120$^{b}$ & 190.37 & -27.40 & 0.03301 & 12.30 \nl 
NGC~3227 &216.99 &55.45 &0.00386 &2.15 \nl
NGC~3516 &133.24 &44.40 &0.00884 & 3.05 \nl
NGC 3783 & 287.46 & 22.95 & 0.00973 &8.50 \nl 
NGC 4051 & 148.88 & 70.09 & 0.00242 & 1.31\nl 
Mkn 766 & 190.68 & 82.27 & 0.01293 & 1.80\nl 
NGC 4593 &297.48 & 57.40 & 0.00831 & 1.97\nl 
MCG-6-30-15 & 313.29& 27.68 & 0.00775&4.06\nl 
IC 4329a & 317.50& 30.92 &0.01605 & 4.55 \nl
Mkn~279 &115.04 &46.86 &0.03045 & 1.64 \nl 
NGC 5548 & 31.96 & 70.50 & 0.01717 & 1.70 \nl 
Mkn 509 & 35.97 & -29.86 & 0.03440 & 4.44 \nl 
NGC~7314 &27.14 &-59.74 &0.00474 & 1.46 \nl
Akn 564 &92.14 & -25.34 & 0.02467 & 6.40 \nl
\enddata
\tablecomments{Columns 2,3 and 4 give 
 the Galactic co-ordinates and the redshift of the source (from  NED). 
Redshift
       was deduced from observations of the 21cm H~{\sc{i}} line where
      possible, since  optical estimates of $z$ may be confused by AGN
 outflow. $^{a}$ Galactic column density from  Elvis, Lockman \& Wilkes 
(1989), except for Mkn
     509 (Murphy \etal 1996).  The Galactic column density towards F9,
           NGC~3227, NGC~3516, NGC~3783, NGC~5548, Mkn 766, NGC~7314 
and  Akn 564 was estimated from
   interpolations from the measurements of  Stark \etal (1992). $^{b}$ 
\thr is also classified as a broad--line radio galaxy (NED).}
\end{deluxetable}

Since there has been a systematic search for \oxysix absorption by hot 
gas in the vicinity of the Galaxy along the sight lines to AGN by 
S03 and since Oxygen is the most abundant element that is detectable 
in the soft X--ray band, we searched the \chandra spectra for evidence of the 
strongest absorption transitions in Oxygen, namely \oxysevenr 1s--2p 
and \oxla (which have oscillator strengths of $f=0.415$ and 
0.696 respectively) in the 
velocity range $cz=0 \pm 1200$ km $\rm{s}^{-1}$ in the local standard of 
rest (LSR). Once we measured the Oxygen absorption profiles, we used the 
extrapolated linear approximation to the curves--of--growth\footnote{
The linear part of the curves--of--growth implies that 
$N_{ion}=1.13 \times 10^{17} EW/ f \lambda^{2}$ 
where $N_{ion}$ is the ionic column density ($\rm{cm}^{-2}$), EW is the 
equivalent width of the absorption feature (in m\AA\ ), 
f is the oscillator strength of the transition and $\lambda$ is in \AA .}  
to obtain a lower limit on the ionic column density ($N_{ion}$), if a 
\emph{lower limit} on the EW of the absorption feature is 
available. Such a lower limit on $N_{ion}$ is valid for any value of the 
velocity width (b) of the absorber. Where no lower limit on 
the EW exists, absorption is not significant (at 90$\%$ 
confidence). However, in this case, for an assumed b value, we can use 
the \emph{upper limit} on the EW (if it exists) to get an upper limit 
on $N_{ion}$. In such cases, we assumed a velocity width of 
$b \sim 100$ km $\rm{s}^{-1}$, since this is about the smallest 
width of a feature that the MEG can resolve although it 
is considerably larger than the average value of 
$<b>= 40 \pm 13$ km $\rm{s}^{-1}$ found by S03 in 
the local hot gas along 59 sight lines. S03 indicate 
that the \oxysix along most 
sight lines is collisionally ionized. This suggests that unresolved 
saturated structure is unlikely in the absorption features. We tested the 
possibility of unresolved saturated structure in the absorption features 
(and thereby double--checked the estimates of ionic column density) by 
deducing the threshold depths of the bound--free \oxyseven and 
\oxyeight absorption edges (0.7393 keV and 0.8714 keV in the 
rest--frame respectively). 

\section{Spectral Fitting}
\label{sec:soft}
  We used XSPEC v.11.2.0 for spectral fitting to the MEG spectra. All
spectral fitting was carried out based on the best--fitting continuum 
models from McKernan \etal (2004). Spectral fitting was carried out in the 
0.5--5~keV energy band, excluding the 2.0--2.5 keV region, which suffers 
from systematics as large as $\sim 20\%$ in the effective area due 
to limitations in the calibration of
                                                   the X-ray telescope
\footnote{http://asc.harvard.edu/udocs/docs/POG/MPOG/node13.html}. 
 We analyzed data binned at $\sim 0.02$\AA, which is approximately the
         MEG FWHM spectral resolution ($0.023$\AA). MEG spectral resolution
corresponds to FWHM velocities of $\sim 280$ and $560 \rm km \
   s^{-1}$ at observed energies of 0.5 and 1.0 keV respectively. We used the
C--statistic (Cash 1976) for finding best--fit model parameters and quote 
90$\%$ confidence, one--parameter errors.

\begin{deluxetable*}{lrrrrrrrr}
\tablecaption{Spectral fitting results for hot gas at $z=0$ \label{tab:2}}
\tablecolumns{9}
\tablewidth{0pt}
\tablehead{
\colhead{Source} & \colhead{EW(\oxysevenr)} &\colhead{EW(\oxla)} &\colhead{$\rm{N}_{\rm{O~\sc{VII}}}$} &\colhead{$\rm{N}_{\rm{O~\sc{VIII}}}$} &\colhead{vel} & \colhead{$\rm{N}_{\rm{O~\sc{VI}}}$$^{a}$}
 &\colhead{vel$^{a}$} &\colhead{ID$^{a,b}$} \nl
{} & \colhead{(eV)($\Delta$C)} & \colhead{(eV)($\Delta$C)} & \colhead{($\rm{cm}^{-2}$)} 
& \colhead{($\rm{cm}^{-2}$)} 
& \colhead{(km $\rm{s}^{-1}$)} & \colhead{($\rm{cm}^{-2}$)} 
& \colhead{(km $\rm{s}^{-1}$)}&\nl
}
\startdata
F9 &$0.79^{+c}_{-0.29} (7.7)$ &$<0.67 (0.5)$ &$>16.10$ &$<16.60$  & $-180 \pm 155$ &$14.33 \pm 0.14$ & $+185^{+90}_{-85}$& MS\nl
3C 120 &$0.60^{+c}_{-0.57}(3.0)$ &$0.94^{+c}_{-0.57}(5.3)$ & $>14.85$  & $>15.73$ & $120\pm 105$ &\ldots &\ldots &()\nl
NGC 3783 &$0.61^{+0.27}_{-0.25}(16.8)$ &$<0.36(1.4)$ &$^{<16.65}_{>15.85}$ & $<15.88$ &$190^{+105}_{-155}$ &\ldots &\ldots &(EPn)\nl
NGC 4051 &$1.63^{+c}_{-0.65}(16.8)$ &$0.85^{+0.28}_{-0.32}(12.1)$ &$>16.45$  & $^{<18.38}_{>15.85}$& $110 \pm 115$ & \ldots& \ldots& (EPn)\nl
NGC 4593 &$<0.74(5.5)$ &$<0.70(2.3)$ &$<17.25$ & $<16.75$&$85^{+175}_{-105}$ & \ldots& \ldots& (EPn)\nl
NGC 5548 &$0.69^{+c}_{-0.44}(5.0)$ &$1.06^{+c}_{-0.48}(9.8)$ &$>15.74$ &$>15.93$ & $-70 \pm 165$ &$<13.84^{e}$ & \ldots &()\nl
Mkn 509 &$<0.57(6.7)$ &$0.58^{+0.28}_{-0.43}(4.3)$ &$<16.75$ &$^{<16.13,d}_{>15.28}$ &$100 \pm 155$ &$14.24\pm 0.10$ &$247^{+67}_{-98}$ &LG\nl
        &\ldots & \ldots & \ldots & \ldots &\ldots &$13.76 \pm 0.19$ &$-143^{+43}_{-37}$ &LG\nl
        & \ldots & \ldots & \ldots & \ldots & \ldots & $13.55 \pm 0.28$&$+152^{+48}_{-37}$ &LG\nl
\enddata

\tablecomments{Inverted Gaussians with redshift fixed at z=0 were added 
to the continuum at the rest-frame energies of \oxysevenr and \oxla. 
The Gaussian model component energies were free to vary up to 
$\pm 1200$ km $\rm{s}^{-1}$ from z=0. The continuum model for 
each AGN was the best--fitting powerlaw or broken powerlaw from 
McKernan \etal (2004). Columns 2 and 3 show the best--fit EW for \oxysevenr 
and \oxla respectively and (in brackets) the improvement in the 
fit--statistic upon the addition of the inverted Gaussian model component 
to the continuum. Columns 4 and 5 show the ionic column densities of 
\oxyseven and \oxyeight respectively as estimated from a curve--of--growth 
analysis as described in the text. Lower limit on ionic column densities 
are valid for all values of $b$. Upper limits 
are valid for an assumed velocity width of $100$ km $\rm{s}^{-1}$ 
(unless specified otherwise) which is approximately the lower bound on 
the instrumental resolution. A choice 
of $b=100$ km $\rm{s}^{-1}$ is larger than the values inferred by 
S03 for local \oxysix absorption features. Column 6 shows 
the weighted mean velocity centroid offset from z=0 (LSR) of the two 
absorption features, except for F9, NGC~3783, NGC~4593 and Mkn~509 where 
this is the offset velocity of the strongest absorption feature. 
Column 7 shows the \oxysix column density 
(where available) along the line of sight to the AGN 
as determined by FUSE (S03). Column 8 shows the velocity of the
z=0 \oxysix absorber components (where available). Velocities are rounded 
to the nearest 5 km $\rm{s}^{-1}$ and a negative value denotes 
blueshift from z=0 (in the LSR). Column 9 shows the 
local structure identified by S03 with the \oxysix absorption along the 
sight lines to the AGN (IDs in brackets are those proposed by us). 
$^{a}$ from S03. $^{b}$ LG=Local Group, MS=Magellanic 
Stream, EPn=Extreme Positive north. $^{c}$ No meaningful upper 
limits because of poor statistics and possible line saturation. 
$^{d}$ Valid for all values of b. $^{e}$ At $3 \sigma$ confidence level. 
} 
\end{deluxetable*}

\begin{deluxetable}{lrr}
\tablecaption{Results from fits to Oxygen absorption edges at 
$z=0$ \label{tab:3}}
\tablecolumns{3}
\tablewidth{0pt}
\tablehead{
\colhead{Source} &\colhead{$\rm{N}_{\rm{O~\sc{VII}}}$} &\colhead{$\rm{N}_{\rm{O~\sc{VIII}}}$} \nl
{} & \colhead{($\rm{cm}^{-2}$)}& \colhead{($\rm{cm}^{-2}$)} \nl
}
\startdata
F9 &$<17.22$ &$<17.46$ \nl
3C 120 & $<17.34$ &$17.87^{+0.20}_{-0.23}$ \nl
NGC 3783 &$<16.09$ &$17.21^{+0.27}_{-0.90}$ \nl
NGC 4051 & $17.56^{+0.18}_{-0.30}$ &$<17.70$  \nl
NGC 4593 & $17.17^{+0.29}_{-1.25}$&$<17.54$\nl
NGC 5548 &$<17.27$&$<17.55$ \nl
Mkn 509 &$<17.23$& $<17.51$ \nl
\enddata

\tablecomments{Columns 2 and 3 
show the column densities of \oxyseven and 
\oxyeight respectively as estimated from depths of the corresponding 
bound--free edges (rest--frame edge threshold energies are 
0.7398 and 0.8714 keV for \oxyseven and \oxyeight 
respectively). The absorption cross--sections at the bound--free 
edge threshold energies are $2.42, 0.99 \times 10^{-19} \rm{cm}^{2}$ for 
\oxyseven and  \oxyeight respectively (Verner \etal 1996). The limit on 
$\rm{N}_{\rm{O~\sc{VIII}}}$ in NGC~3783 does not agree with the results from 
Table~\ref{tab:2}. However NGC~3783 has a very complicated continuum 
due to strong absorption intrinsic to the AGN (see e.g. McKernan \etal 2004 
for details).}
\end{deluxetable}

We proceeded to fit the MEG spectra for the Oxygen absorption 
transitions by adding an inverted Gaussian model 
component to the best--fitting continuum models detailed in 
McKernan \etal (2004). We fixed the redshift of the Gaussian 
components at $z=0$ and allowed the rest--energy of the component to vary 
by $\pm 1200$ km $\rm{s}^{-1}$ from the rest--frame energies of \oxysevenr 
(0.5740keV) and \oxla (0.6536keV) respectively. The allowed velocity range 
is identical to that used by S03 in their search for 
\oxysix absorption in the vicinity of the Milky Way. In fitting the 
bound--free absorption edges, we fixed the edge energies at the rest--frame 
values (0.7393 and 0.8714 keV for \oxyseven and \oxyeight 
respectively) and we fixed  the redshift at $z=0$.

\section{Results}
\label{sec:results}
Of the fifteen AGN in our sample, we find that seven AGN exhibit an 
improvement in the fit--statistic at $>90\%$ confidence ($\Delta C \geq 6.3$ 
for three additional parameters) when we fit the spectra with 
inverted Gaussian model components corresponding to highly 
ionized Oxygen absorption features (\oxysevenr or \oxla or both) 
at $z=0$. Table~\ref{tab:2} shows the spectral 
fitting results for these seven AGN, including the equivalent 
widths (EW) of the 
absorption features, the velocity offset from LSR of the Gaussian centroid 
and limits on the column densities of highly ionized Oxygen as 
estimated from a curve--of--growth analysis as outlined in \S\ref{sec:obs}. 
Also listed in Table~\ref{tab:2} are the column density and offset 
velocity from LSR of \oxysix along the sight lines (from S03) 
where available. The last column of Table~\ref{tab:2} lists the local 
structure 
associated by S03 with the sight lines to each AGN (where 
available). We could not obtain meaningful 
upper limits from Gaussian fits to the EW of six of the sixteen 
absorption features listed in Table~\ref{tab:2}. This is most likely due 
to poor statistics, but it could also indicate line saturation.

\oxysevenr absorption due to local hot gas is present in the spectra of 
NGC~3783 and NGC~4051 at $>3 \sigma$ significance 
($\Delta C \geq 14.2$ for 3 additional parameters) and in the spectra of 
F9 and Mkn~509 at $>90\%$ confidence 
($\Delta C \geq 6.3$ for 3 additional parameters). \oxyeight absorption due 
to local gas is present only in the spectra of NGC~5548 and 
NGC~4051 at $>90\%$ confidence. Although both of the absorption signatures 
towards 3C~120 and NGC~4593 are detected at $<90\%$ confidence, we 
include these results in Table~\ref{tab:2} because the detection of 
\emph{both} local 
absorption signatures corresponds to a confidence level $>90\%$. 
Of the fifteen AGN spectra in our sample only one (NGC~4051) exhibits 
absorption due to \emph{both} local \oxysevenr \emph{and} local \oxla --
at $>3 \sigma$ and $>99\%$ confidence ($\Delta C \geq 14.2$ and 
$\geq 11.3$ for 3 additional parameters respectively). 

Table~\ref{tab:3} shows the results from fitting \oxyseven and \oxyeight 
absorption edges at $z=0$ to the spectra of AGN in Table~\ref{tab:2}. 
In general, the absorption edge 
fits in Table~\ref{tab:3} provide upper limits to the \oxyseven and 
\oxyeight column densities of the hot, local gas and agree with the 
limits on $\rm{N}_{\rm{O~vii}}$, $\rm{N}_{\rm{O~viii}}$ from 
Table~\ref{tab:2}. 
The exception is the limit on $\rm{N}_{\rm{O~viii}}$ in NGC~3783, 
which does not agree with Table~\ref{tab:2}. However NGC~3783 has a 
very complicated continuum due to strong absorption intrinsic to 
the AGN (see e.g. McKernan \etal 2004 for details).

\begin{figure*}
\epsscale{0.8}
\plotone{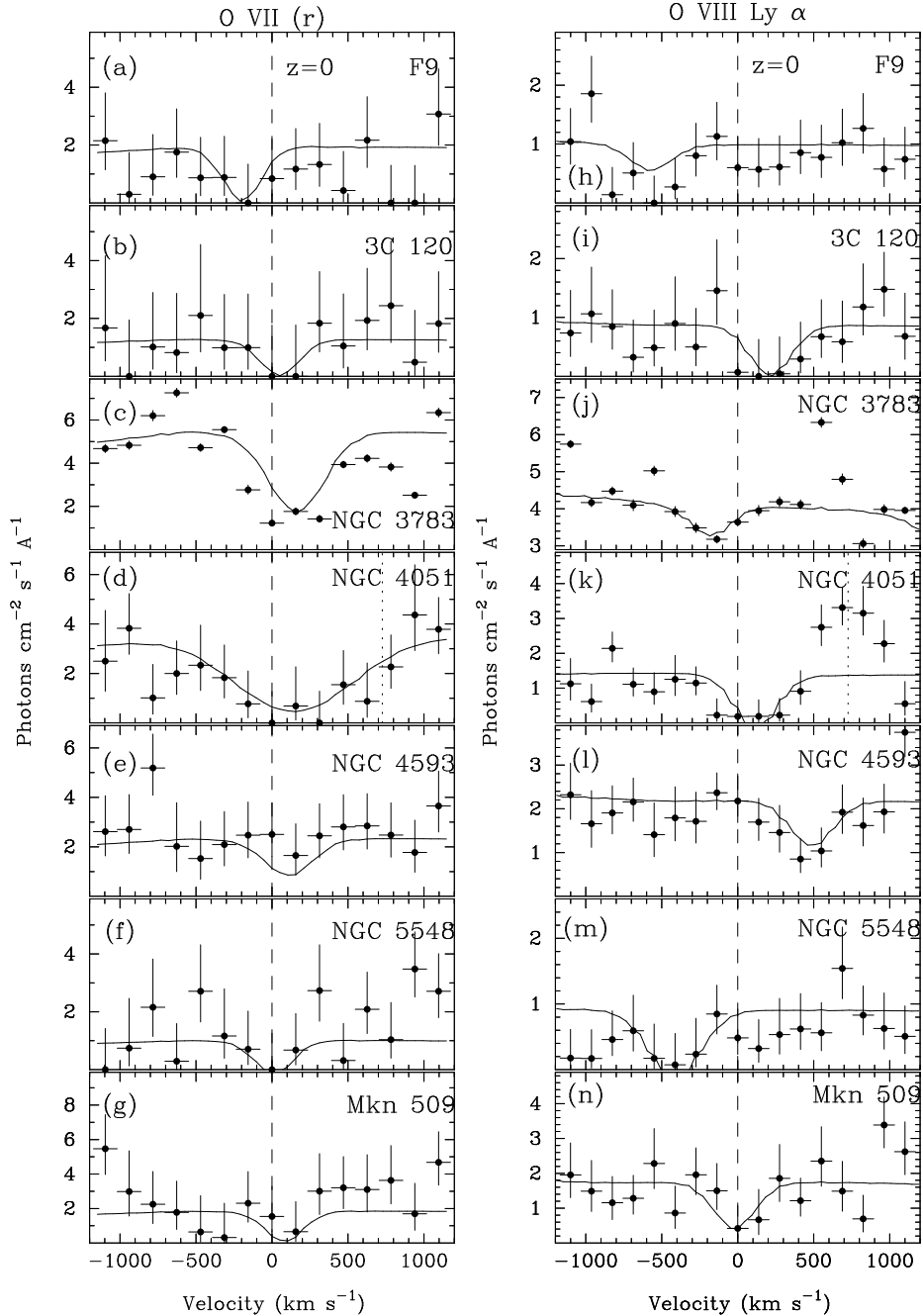}
\caption{Velocity profiles from combined $\pm$1 order \chandra MEG data from 
the AGN in Table~\ref{tab:2}, centered on the LSR 
\oxysevenr transition energy (0.57396 keV) in the left column panels and 
on the \oxla transition energy (0.65362 keV) in the right column panels. 
A positive velocity indicates a redshift relative to these energies. The 
velocity spectra data have been uniformly binned at 0.3eV which is 
approximately the limit of the MEG resolution. Vertical dashed lines 
at 0 km $\rm{s}^{-1}$, correspond to the location 
of the \oxysevenr transition energy (left panels) or the 
\oxla transition energy 
(right panels) in the LSR. Vertical dotted lines in panels (d) and (k) 
indicate the rest--frame of NGC~4051. The remaining AGN rest--energies lie
 outside the $\pm 1200$ km $\rm{s}^{-1}$ range of these panels. Note 
that in NGC~4051, the broad absorption features indicate that there may 
be a blend of intrinsic warm absorption in the AGN (outflowing at 
$\sim 600$ km $\rm{s}^{-1}$) and local absorption. 
 Superposed is the best--fit inverted Gaussian absorption line model 
(from Table~\ref{tab:2}) and continuum (horizontal solid line). The models 
are calculated from the best--fit Gaussian model in Table~2 since most of the 
results in Table~\ref{tab:2} are independent of the value of b 
($=\sqrt{2} \sigma$).
\label{fig:vel_profiles}
}
\end{figure*}

Fig.~\ref{fig:vel_profiles} is a multipanel plot showing velocity profiles 
from the AGN in Table~\ref{tab:2}. The profiles are centered on the 
\oxysevenr transition energy in the LSR (0.5740 keV) and the 
\oxla transition energy (0.6536 keV) respectively (both energies are 
denoted by vertical dashed lines). Superposed is the best--fit 
inverted Gaussian absorption line model 
(from Table~\ref{tab:2}) and continuum (horizontal solid line). 
The vertical dashed line in 
Fig.~\ref{fig:vel_profiles}(d,k) at $+725$ km $\rm{s}^{-1}$ denotes the 
redshift velocity (cz) of NGC~4051. Note that absorption towards this AGN 
could correspond either to absorption by gas in the vicinity of our Galaxy 
or intrinsic absorption in the AGN at an outflow velocity of 
$615 \pm 115$ km $\rm{s}^{-1}$ (see discussion in \S\ref{sec:outflow}). 
There is a hint of additional absorption at $\sim -400$ km $\rm{s}^{-1}$ 
towards Mkn~509 (Fig.~\ref{fig:vel_profiles}(g,n)). According to S03 
there are three 
UV absorption components due to LG gas along this sight line spanning 
$\sim 550$ km $\rm{s}^{-1}$ in offset velocity. Therefore, we might 
expect broad absorption profiles from a blend of several narrow 
components along the sight line to Mkn~509. In fact, a considerably 
broader and deeper \oxysevenr profile can be fit to the Mkn~509 data, 
but the required width of the resulting feature is greater than the 
width of the velocity 
range that we are studying here (2,400 km $\rm{s}^{-1}$. Collins \etal (2004) 
suggest that broad, high velocity \oxysix absorption along this 
sight line arises from shock 
ionization at bow shock interfaces produced from infalling high velocity 
clouds (HVCs). 

\begin{deluxetable}{lrr}
\tablecaption{Null results for hot gas at $z=0$ \label{tab:4}}
\tablecolumns{3}
\tablewidth{0pt}
\tablehead{
\colhead{Source} &\colhead{$\rm{N}_{\rm{O~\sc{VII}}}$} 
&\colhead{$\rm{N}_{\rm{O~\sc{VIII}}}$} \nl
{} & \colhead{($\rm{cm}^{-2}$)}& \colhead{($\rm{cm}^{-2}$)} \nl
}
\startdata
Mkn 766 &$<17.85$ &$<16.28$ \nl
MCG-6-30-15 & $<16.90$ &$<16.18$ \nl
IC 4329A &$<17.85$ &$<16.13$ \nl
Mkn 279 & $<17.65$ &$<18.58$  \nl
Akn 564 & $<16.10$&$<15.58$\nl
\enddata

\tablecomments{as for Table~\ref{tab:2}. The spectra of NGC~3227, NGC~3516 
and NGC~7314 were too strongly absorbed at $<0.7$ keV to obtain limits on
$\rm{N}_{\rm{O~\sc{VII}}}$ and $\rm{N}_{\rm{O~\sc{VIII}}}$.}
\end{deluxetable}

Eight AGN from the fifteen listed in Table~\ref{tab:1} did not exhibit an 
improvement in the fit--statistic when their spectra were fit with 
\oxysevenr and \oxla absorption features. The spectra of three 
of these eight AGN, namely NGC~3227, NGC~3516 and NGC~7314 were so 
strongly absorbed at energies $\leq 0.7$ keV that it was not possible 
to obtain limits on the column densities of $\rm{N}_{\rm{O~vii}}$ and 
$\rm{N}_{\rm{O~viii}}$ along these sight lines. Table~\ref{tab:4} lists the 
upper limits of $\rm{N}_{\rm{O~vii}}$ and $\rm{N}_{\rm{O~viii}}$ along the 
sight lines 
to the remaining five AGN. Of these five AGN, only Akn~564 has a sight line 
that coincides with the S03 study, and exhibits \oxysix absorption 
associated by S03 with the local MS extension.

\section{Comparison with \oxysix absorption results}
\label{sec:compare}

\begin{figure*}
\epsscale{1.0}
\plotone{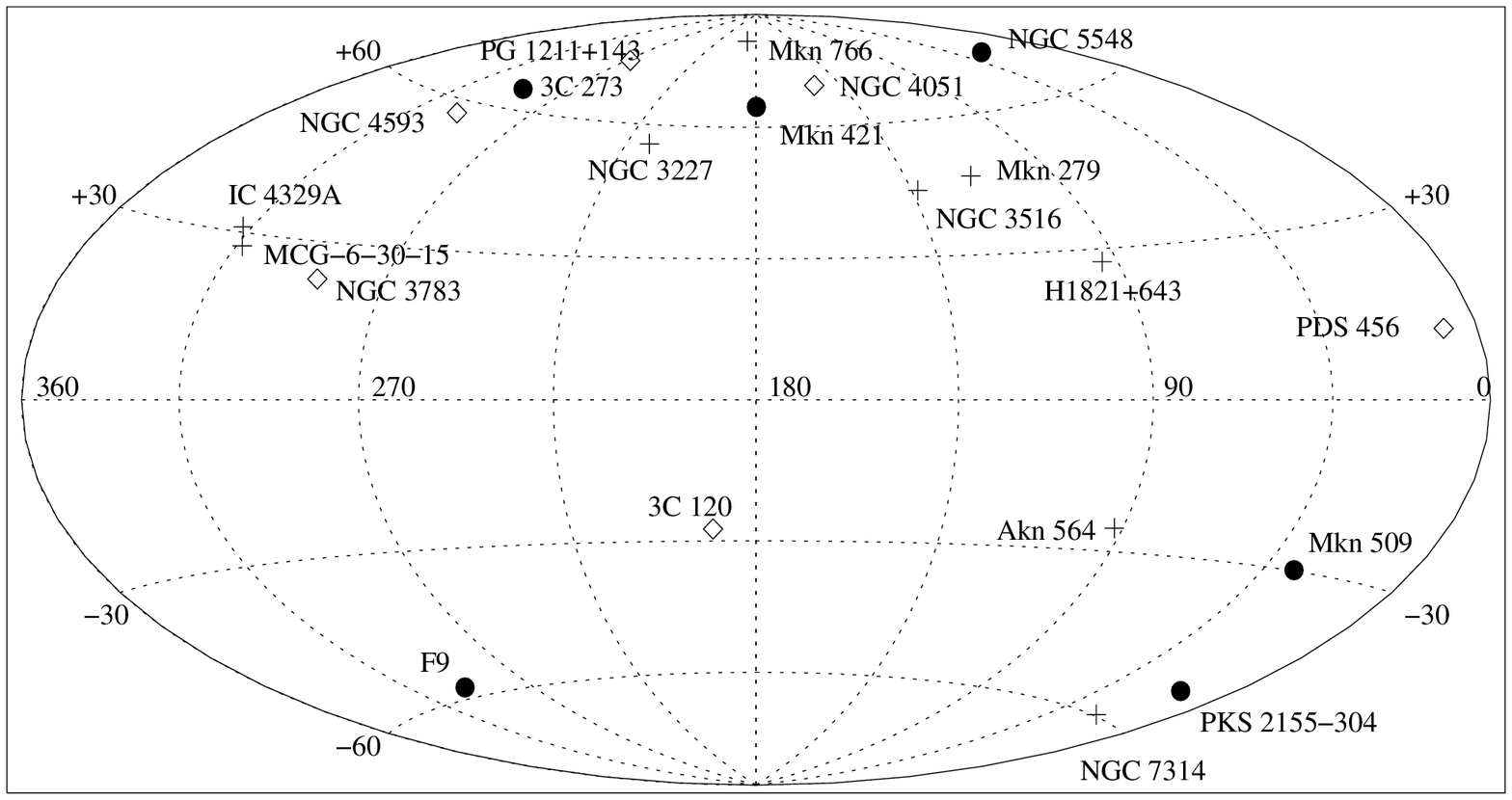}
\caption{All-sky Hammer-Aitoff projection of the sight lines to the 
15 AGN in this study, plus sight lines for six additional AGN in the 
literature. The additional AGN are Mkn 421 (Cagnoni 2002; 
Rasmussen \etal 2003), 
PKS 2155--304 
(Nicastro \etal 2002), 3C 273 (Fang \etal 2003), PG~1211$+$143 
(Pounds \etal 2003), PDS~456 (Reeves \etal 2003) and H 1821+643 (Mathur 
\etal 2003). In this projection, the Galactic anticenter is at the 
center of the figure and Galactic longitude increases to the left. 
Plus symbols indicate nondetection of local, hot gas. Open, diamond symbols 
indicate detection of local, hot gas along the sight lines at $>90\%$ 
confidence but with no corresponding \fuse observations. Filled--in 
symbols denote detections of local, hot gas along the sight lines \emph{and} 
corresponding \fuse observations of \oxysix. PG~1211$+$143 and PDS~456 are 
included because the apparent intrinsic absorption in their spectra has an 
outflow velocity which coincides with our LSR.\label{fig:allsky_map}
}
\end{figure*}

S03 searched for \oxysix ($\lambda 1031.926$ \AA) 
absorption along 102 sight lines, at high Galactic latitudes of 
$|b| \geq 30^{\circ}$ towards UV bright AGN/QSOs. Some 59 of the 102 sight 
lines 
revealed a total of 84 high velocity ($|v|>100$ km $\rm{s}^{-1}$ in the 
LSR) \oxysix absorption components at 
$\geq 3 \sigma$ confidence level. Three of these 59 sight lines coincide with 
sight lines in our sample (see Table~\ref{tab:2}). The sight lines to 
the fifteen AGN in our sample plus six AGN in the literature towards 
which WHIGM at $z=0$ has been detected or searched for, are shown 
in Hammer--Aitoff projection in Figure~\ref{fig:allsky_map}. Crosses 
indicate a lack of highly ionized local gas (at $90\%$ confidence). 
Open, unfilled symbols indicate detections (at $>90\%$ confidence) 
of highly ionized local gas with no corresponding 
\fuse sight lines. Filled--in symbols indicate sight lines along which 
highly ionized 
Oxygen has been detected (at $>90\%$ confidence), with corresponding 
\fuse observations of 
\oxysix absorption. It is useful to compare this 
plot with similar projections in S03 
(see e.g. their Fig.~6). Whilst the number of points in 
Fig.~\ref{fig:allsky_map} is 
small, we can begin to associate very highly ionized absorption along 
some of these sight lines with the hot structures discussed by 
S03. 

The average value of the 
\oxysix column density ($\rm{N}_{\rm{O~vi}}$) detected by S03 was 
$\langle \log \rm{N}_{\rm{O~vi}} \rangle = 13.95 \pm 0.34$, which is 
considerably lower than the average column of 
$\langle \log \rm{N}_{\rm{O~vi}} \rangle = 14.38 \pm 0.18$ in the Galactic 
thick disk/halo. By comparison, along the sight lines in 
our sample 
$\langle \log \rm{N}_{\rm{O~vii}} \rangle,\langle \log \rm{N}_{\rm{O~viii}} 
\rangle$ seem to 
lie in the range $\sim 16-17$ (see Table~\ref{tab:2}). The highest 
values of $\rm{N}_{\rm{O~vi}}$ found by S03 
are associated by them with known local structures such 
as the MS, Complex C and Extreme Positive North (EPn). The 
highest values of $\rm{N}_{\rm{O~vii}}$ and $\rm{N}_{\rm{O~viii}}$ from our 
Table~\ref{tab:2} may also be associated with local structure (EPn and MS). 

\oxysix in low redshift IGM is far more likely to be 
collisionally ionized than photoionized (Heckman \etal 2002; S03). 
Therefore, if we assume that the hot gas in the 
vicinity of the Galaxy is in collisional ionization equilibrium (CIE), and
that the \oxysix and \oxyeight absorption occurs in the same gas, we can 
establish temperature constraints on the gas. Sutherland \& Dopita 
(1993) calculate $\rm{N}_{\rm{O~vii}}$/$\rm{N}_{\rm{O~vi}}$ and 
$\rm{N}_{\rm{O~viii}}$/$\rm{N}_{\rm{O~vi}}$ for 
gas in CIE and we can use these values to constrain the temperature 
of the local gas towards several of the AGN in our sample. In the 
case of the F9 sight line, CIE implies $10^{5.75}T<10^{6.35}$K. For the 
gas along the sight line to Mkn~509, CIE implies 
$T< 10^{6.20}$ K, although Collins \etal (2004) suggest that 
non-equilibrium ionization occurs along this sight line, in 
which case CIE is ruled out. Likewise, if the hot gas along 
the sight line to NGC~5548 and Akn~564 is in CIE, then  $T>10^{5.80}$K and 
$T<10^{6.10}$K for the NGC~5548 and Akn~564 sight lines respectively. 

S03 find that \oxysix absorption features at $z=0$ have a 
mean centroid velocity relative to LSR of $-33 \pm 207$ km $\rm{s}^{-1}$. 
The mean centroid velocity of the features listed in Table~\ref{tab:2} is 
$70 \pm 25$ km $\rm{s}^{-1}$. S03 found that the centroids 
of the \oxysix features are generally significantly redshifted at Galactic 
longitudes $180^{\circ} < l< 360^{\circ}$ and significantly blueshifted at 
Galactic longitudes $0^{\circ} < l< 180^{\circ}$. The limited size of our 
sample and the limited velocity resolution of \chandra (MEG has a FWHM 
resolution of $\sim 280$ km $\rm{s}^{-1}$ at energies around \oxysevenr) 
mean that we cannot establish a similar pattern in our data. 
S03 also found that the average high velocity feature 
width is $<b>= 40 \pm 13$ km $\rm{s}^{-1}$. By comparison, we have 
\emph{assumed} a turbulent velocity of $b \sim 100$ km $\rm{s}^{-1}$ 
where necessary to establish limits on ionic column density. Around half the 
features 
shown in Fig.~\ref{fig:vel_profiles} and listed in Table~\ref{tab:2} may 
be saturated, suggesting that the actual turbulent velocity of 
the gas may be $<100$ km $\rm{s}^{-1}$, however we are limited by the 
MEG velocity resolution and (generally) the low signal to noise of the data 
in this energy range.

The Local Group (LG) features discussed by S03, include 
the Local Group (LG) itself and extreme 
positive north (EPn). The average offset velocity of the centroid 
of the absorption features that S03 identify with the 
Local Group itself is $-55 \pm 212$ km $\rm{s}^{-1}$ in the LSR. 
LG features have \emph{no} obvious corresponding H~{\sc i} 
21 cm emission at similar velocities directly along the sight lines. All 
of the LG clouds are located in the southern Galactic hemisphere, mostly 
in the $35^{\circ} < l < 140^{\circ}$ Galactic longitude range (similar to 
the range in Galactic longitude of the MS extension absorption 
features). Four of 
the seven sight lines listed in Table~\ref{tab:2}, namely NGC~3783, NGC~4051, 
NGC~4593 
(all EPn) and Mkn 509 (LG) can likewise be identified with the Local Group.

The sight line to Mkn~509 passes through two high velocity 
clouds (HVCs) detected in absorption in C~{\sc iv} with velocity centroids at 
$-283, -228$ km $\rm{s}^{-1}$. The HVCs have ionization properties 
consistent with clouds irradiated by local extragalactic background 
radiation according 
to Sembach \etal (1999), however the amount of N O~{\sc{vi}} inferred along 
the sight line suggests that this is not the only source of ionization. 
S03 and Collins \etal (2004) suggest that these clouds interact with a hot 
Galactic corona or a low density LG medium (the $z=0$ WHIGM). 
As noted above, the \oxysevenr feature in Fig.~\ref{fig:vel_profiles}
(f) can be fit with a broader inverted Gaussian profile than in 
Table~\ref{tab:2} since the line shape is poorly constrained. The 
\oxyeight feature along this sight line appears 
much weaker however, which suggests an upper limit to the temperature of 
the local gas along this sight line.

In the southern Galactic hemisphere, the sight line 
to F9 passes through the MS and the sight line to Akn~564 includes the 
Magellanic Stream extension (MSe), but this may also be LG (S03). 
As discussed above, 
if the UV and X--ray absorbing hot gas along the 
sight lines to F9 is the same, and if this gas is in CIE, the temperature 
in the 
MS is constrained to lie in the range $10^{5.75}<T<10^{6.35}$K. We detect 
little or no 
\oxyseven or \oxyeight absorption towards Akn~564, which suggests that 
the temperature of the hot-phase of the MS extension is lower than that in 
the MS. Also in the 
southern Galactic hemisphere, the sight line to 3C~120 does not 
coincide with 
any large scale local structure. Hot, local aborption along this sight line 
may correspond to the local WHIGM. In the northern Galactic 
hemisphere, the sight lines to
NGC~3783, NGC~4593 and NGC~4051 overlap with the EPn structure as 
described by S03, however we cannot obtain any 
temperature constraints along these sight lines, since these sources were not 
part of the S03 study. The sight line to NGC~5548 is not 
associated with any known local structure. Among the AGN not contained in our sample but indicated in Fig.~\ref{fig:allsky_map}, 
we note that the sight line to PKS~2155-304 (Nicastro \etal 2002; 
Fang \etal 2003) passes through the two HVCs that also lie 
along the sight lines to Mkn~509. The sight line to H~1821+643 
(Mathur \etal 2003) 
passes through the Outer Spiral Arm and the sight lines to both 3C~273 
(Fang \etal 2003) and Mkn~421 (Cagnoni 2002; Rasmussen \etal 2003) passes through EPn. 
Of course there are Galactic sources that could contribute highly 
ionized Oxygen along the sight lines to several of the AGN. The Local Bubble 
can contribute at most only $log$ N O~{\sc{vii}} $\leq 15.48$ and 
Radio Loops, 
possibly corresponding to Supernova Remnants (SNR), along the sight line to 
3C~273 could contribute $\log$ N O~{\sc{viii}} $\leq 15.90$ 
(Fang \etal 2003). 
The magnitudes of $\rm{N}_{\rm{O~vii}}$ and $\rm{N}_{\rm{O~viii}}$ in 
Table~\ref{tab:2} suggest that 
sources within the Milky Way contribute at most a small fraction of 
the \oxyseven and \oxyeight column density along the sight lines to the AGN 
in our sample.

\section{Is apparent AGN outflow actually due to local absorption?}
\label{sec:outflow}
Some components of the X-ray absorption which is apparently 
intrinsic to an AGN, may actually be due to absorption by hot, local gas. 
For example, a low velocity warm 
absorber component has been detected in NGC~4051 at a blueshift 
with respect to the AGN rest--frame of $-575 \pm 175$ km $\rm{s}^{-1}$ 
(Collinge \etal 2001) or $-645 \pm 130$ km $\rm{s}^{-1}$ 
(McKernan \etal 2004). Since NGC~4051 is at 
$cz=726 \pm 15$ km $\rm{s}^{-1}$, the apparent outflow from this source is 
equivalent to absorption at a redshift of $cz=+150 \pm 175$ km $\rm{s}^{-1}$ 
(using the Collinge \etal 2001 estimate) or $cz=+75 \pm 130$ km 
$\rm{s}^{-1}$ (using the McKernan \etal 2004 estimate) with respect to 
$z=0$. This is certainly consistent with an origin in hot, local gas. 

\begin{figure}
\epsscale{1.0}
\plotone{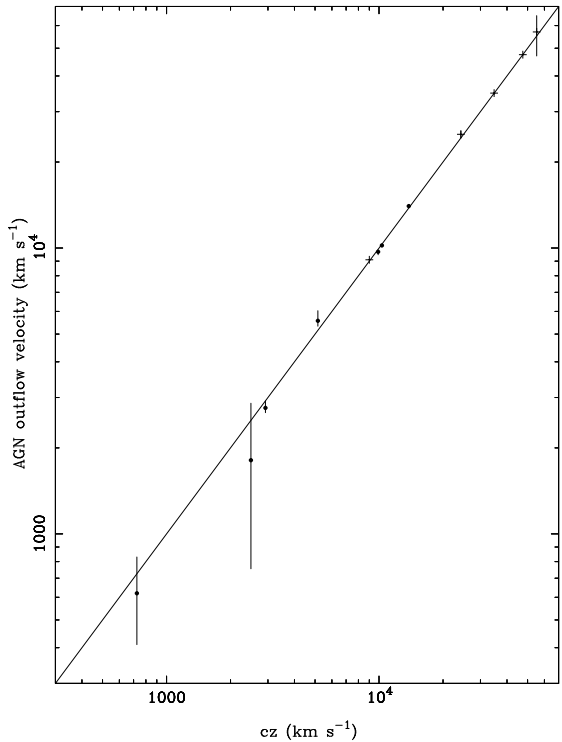}
\caption{Plot of offset velocity of soft X-ray absorption features 
relative to AGN rest--frame versus cosmological recession velocity (cz) of 
AGN. Filled--in symbols denote AGN from Table~\ref{tab:2}. 
Plus symbols denote AGN observations in the literature, 
including Mkn 421 (Cagnoni 2002; Rasmussen \etal 2003), 
PKS 2155--304 (Nicastro 
\etal 2002), 3C 273 (Fang \etal 2003), PDS 456 (Reeves \etal 2004) 
and PG~1211+143 (Pounds \etal 2003). The straight line corresponds to the 
LSR. Most of the 
errors on the velocity measurements are very small (few hundred km 
$\rm{s}^{-1}$) relative to the actual offset velocity (few 
$\times 10^{3-4}$ km $\rm{s}^{-1}$). The apparent AGN 
outflow detected in PDS~456 (rightmost point) by 
Reeves \etal (2004) and in PG~1211+143 (fourth point from right) by 
Pounds \etal (2004) coincide remarkably with the redshifted 
velocity (cz) of the AGN. \label{fig:cz}}
\end{figure}

Fig.~\ref{fig:cz} shows the offset velocity of absorption features 
relative to systemic AGN velocity versus cosmological recession velocity 
(cz) of the AGN. Filled--in symbols denote AGN listed in
 Table~\ref{tab:2} and plus symbols denote AGN discussed in 
the literature. Of the AGN in the literature, the absorption in the spectra 
of Mkn 421, 3C 273 and PKS~2155--304 has been interpreted as due to hot, 
local gas and the absorption in the spectra of PG~1211+143 and PDS~456 has 
been interpreted as due to mildly relativistic outflows from the AGN. 
However, it is remarkable that the absorption detected in the QSO 
PDS~456 (rightmost point) by Reeves \etal (2003) and in the QSO PG~1211+143 
(fourth point from right) by Pounds \etal (2003) should coincide so closely
to absorption at $z=0$. Pounds \etal (2003) detect an
 ionized outflow of $\sim 24,000$ km $\rm{s}^{-1}$ in the \xmm 
spectrum of PG~1211+143 with respect to the QSO rest--frame. However, 
this quasar lies at $cz=24,253 \pm 150$ km $\rm{s}^{-1}$. Similarly, 
Reeves \etal (2003) find that the \xmm spectrum of PDS~456 requires a 
warm absorber with a very high outflow velocity ($47^{+35}_{-9} 
\times 10^{3}$ km $\rm{s}^{-1}$ from the EPIC-PN instrument and 
$57^{+8}_{-10} \times 10^{3}$ km $\rm{s}^{-1}$ from the RGS 
instrument). However the redshift of PDS~456 corresponds to a 
velocity offset of $cz=55,200 \pm 300$ km $\rm{s}^{-1}$. The correlation 
in Fig.~\ref{fig:cz} is impressive, which suggests that some portion 
of the apparent outflows in these two AGN (and possibly in other AGN) may 
be due to absorption by hot, local gas. However, a major problem with a 
local origin of the absorption features in these two QSOs is the 
implied column 
density of highly ionized Fe measured by Pounds \etal (2003) and 
Reeves \etal (2003) which is far too large to be consistent with 
present theories of the Galactic environment and led Pounds \etal (2003) 
and Reeves \etal (2003) to reject a local origin. On the other hand, 
the ionic column densities inferred from a curve--of--growth analysis of 
the soft X-ray absorption features in the spectra of PG~1211+143 and 
PDS~456 seem to be consistent with the column densities that we find 
along the sight lines discussed in this paper. Therefore a kinematic 
agreement between low energy absorption signatures and Fe-K band
 absorption signatures does not \emph{ipso facto} support an intrinsic 
AGN origin for the soft X-ray absorption features. Nevertheless, in the 
absence of a change in our understanding of the local Galactic environment, 
it seems that only a portion of the warm absorption apparently intrinsic 
to PDS~456 and PG~1211+143 could also be due to hot, local gas. 

\section{Conclusions}
\label{sec:conclusions}
We have assembled a small sample of type~I AGN observed with the
high resolution X-ray gratings on board \emph{Chandra} and have applied 
a uniform analysis to detect soft X-ray absorption by hot gas in the 
vicinity of our Galaxy. Around half of the sight lines in our sample 
exhibit local ($z=0$) 
\oxysevenr or \oxyeight absorption (or both) at 
$\geq 90 \%$ confidence. We identify the 
absorption features with hot gas in local structures, following 
the discussion in S03 of local \oxysix absorption. 
Several sight lines in our sample coincide with sight lines in the 
S03 
study and so it is possible to derive constraints on the 
temperature of the local hot gas, assuming CIE.

In the northern Galactic hemisphere, we detect strong absorption by local 
\oxysevenr along the sight lines to NGC~3783 and 
NGC~4051 (which also exhibits \oxyeight absorption). These sight lines 
(and that towards NGC~4593) overlap with the EPn local structure discussed 
by S03. The sight line towards NGC~5548 shows local \oxyseven and 
\oxyeight absorption implying that the gas temperature is $T>10^{6.2}$K 
if the gas is in CIE and we use the S03 upper limit on $\rm{N}_{\rm{O~vi}}$ 
along this sight line. In the southern Galactic hemisphere, we can 
identify absorption along the sight line to F9 with the Magellanic Stream 
(MS). \oxysix is detected towards F9 by 
S03 and if we associate our detection of 
\oxyseven absorption with the same gas, 
then $10^{5.75}<T<10^{6.35}$K if the gas is in CIE. There is a hint of broad 
\oxyseven absorption along the sight line towards Mkn~509. S03 and 
Collins \etal (2004) 
suggest that Local Group clouds along this sight line interact 
with a hot Galactic corona or a low density LG medium 
(possibly the $z=0$ WHIGM), if so CIE is unlikely. The 
sight line to 3C~120 does not 
appear to be associated either with local structure or UV absorbing 
gas (S03), and so absorption may result from local WHIGM. The 
local gas along the sight line to Akn~564 has $T<10^{6.10}$K if 
it is in CIE. A very important point to note is that at least a portion of 
the apparent warm, high--velocity outflows detected in PG~1211+143
 (Pounds \etal 2003) and PDS~456 (Reeves \etal 2003) could actually 
correspond to absorption by hot gas 
in the vicinity of our Galaxy. The low--velocity warm absorber in 
NGC~4051 (Collinge \etal 2001; McKernan \etal 2004) could also 
correspond to local absorption. The correlation in Fig.~\ref{fig:cz} is 
so strong that it seems likely that absorption by hot, local gas may 
account for at least a portion of the apparent outflows in other AGN. 

This is the first attempt to systematically investigate the X--ray 
absorbing hot gas in the vicinity of the Galaxy. We have detected 
multiple absorption signatures along several sight lines through this gas 
at confidence levels ranging from $>90\%$ to $>3\sigma$. Some 
of the gas is associated with known local structures and some may be 
associated with the local WHIGM. Obviously the size of our 
present sample is the main limitation on our investigation of the local, 
hot gas. We need many more sight lines through the local $z=0$ hot gas to 
begin to constrain and genuinely map the WHIGM and the hot phase of 
local structures.

We gratefully acknowledge support from NSF grant AST0205990 (BM,CSR) and NASA 
grants G01 2102X, G02 3133X, AR4-5009X issued by CXC operated by SAO under 
NASA contract NAS8-39073 (TY). Thanks to Smita Mathur for useful 
discussions. We made use of the HEASARC on-line data archive services, 
supported by NASA/GSFC and also of the NASA/IPAC Extragalactic Database 
(NED), operated by the Jet Propulsion Laboratory, CalTech, under contract 
with NASA. Thanks to the \chandra instrument and operations teams for 
making the observations possible.

\end{document}